\documentclass[11pt,aps,prd,showpacs,showkeys,superscriptaddress,preprintnumbers,amsmath,amssymb,nofootinbib]{revtex4-1}

\pdfoutput=1 

\usepackage{graphicx}
\usepackage{dcolumn}
\usepackage{bm}
\usepackage{color}
\usepackage{epsfig}
\usepackage{amssymb}
\usepackage{amsmath}
\usepackage{amsfonts}
\usepackage{nicefrac}
\usepackage{physics}

\usepackage[toc,page]{appendix}
\usepackage{natbib}
\usepackage{amsmath,amssymb,enumitem}
\usepackage{fontenc}[T1]
\usepackage[utf8]{inputenc}
\usepackage{multirow}
\usepackage{placeins}
\usepackage{slashed}
\usepackage{xspace}
\usepackage[usenames, dvipsnames]{xcolor}
\usepackage{url}
\usepackage[linktocpage=true]{hyperref}
\hypersetup{
    colorlinks = true,
    linkcolor = blue,
    citecolor = blue,
    urlcolor = blue}
    
\usepackage{amsmath}
\usepackage{siunitx}
\usepackage{textcomp}

\usepackage[caption=false]{subfig}
\usepackage{pgfplots}
\usepackage{tikz}

\usetikzlibrary{decorations.pathmorphing}
\usetikzlibrary{decorations.markings}
\usepgfplotslibrary{external,statistics}
\usepackage{mathtools}

\newcommand{\be}{\begin{equation}}
\newcommand{\ee}{\end{equation}}

\makeatletter

\def\ak{\@ifstar\@@ak\@ak}
\newcommand{\@ak}[1]{\textcolor{ForestGreen}{[\textbf{AK:} #1]}}
\newcommand{\@@ak}[1]{\textcolor{ForestGreen}{#1}}
\makeatother

\allowdisplaybreaks
\begin{document}

\title{Multipole Moments of Double Heavy $J^P = \frac{3}{2}^+$ Baryons}

\author{T.~M.~Aliev}
\email[Electronic address:~]{taliev@metu.edu.tr}
\affiliation{Physics Department, Middle East Technical University, 06531 Ankara, Turkey}
\author{E.~Askan}
\email[Electronic address:~]{easkan@metu.edu.tr}
\affiliation{Physics Department, Middle East Technical University, 06531 Ankara, Turkey}
\author{A.~Ozpineci}
\email[Electronic address:~]{ozpineci@metu.edu.tr}
\affiliation{Physics Department, Middle East Technical University, 06531 Ankara, Turkey}


\begin{abstract}
\vspace{1cm}
    In the present study, we calculate the multipole moments of spin-3/2 doubly heavy baryons within the light cone QCD sum rules. We compare our results on magnetic dipole moments with results existing in the literature. The results obtained in the present work may be useful for a deeper understanding of the properties of doubly heavy baryons as well as in the analysis of their strong and electromagnetic decays.
\end{abstract}


\preprint{}

\maketitle

\section{Introduction}
    The quark model predicts the existence of baryons containing two heavy quarks. These states represent an attractive platform to study the heavy quark symmetry and chiral dynamics\cite{Gross:2022hyw}. In dependence of the strangeness, doubly heavy baryons are divided into two families, $\Xi_{QQq}$ and $\Omega_{QQs}$ where $\Xi_{QQq}$ contains $u$ or $d$ quarks. Many of the doubly heavy baryons have not yet been observed in experiments\cite{Chen:2022asf}. 
The first observation of a doubly heavy baryon $\Xi_{ccd}^+(3520)$ was announced by the SELEX Collaboration \cite{SELEX:2002wqn} and confirmed by the same collaboration \cite{ocherashvili2005a}.
In 2017, the LHCb collaboration observed $\Xi_{ccu}^{++}$ in the $\Lambda_c^+K^-\pi^+\pi^-$ channel with the mass $3621$ MeV \cite{LHCb:2017iph}. This state was also confirmed by the LHCb collaboration in the $\Xi_{cc}^{++}\rightarrow\Xi_c^+\pi^+$ decay channel \cite{LHCb:2018pcs}. Other states are still not experimentally observed, despite intensive experimental efforts.
These discoveries inspired many theoretical works for the study of the mass, lifetimes, and strong coupling constants with light mesons as well as weak, electromagnetic, and strong decays of the doubly heavy baryons\cite{Aliev:2009pd, Aliev:2020aon, Aliev:2021hqq, aliev2023, Alrebdi:2020rev, Azizi:2020zin, Cheng:2020wmk, Dhir:2018twm, Gerasimov:2019jwp, Gutsche:2018msz, Gutsche:2019iac, Gutsche:2019wgu, Han:2021gkl, Hu:2017dzi, Hu:2020mxk, Jiang:2018oak, Ke:2019lcf, Li:2017pxa, Li:2018epz, Li:2020qrh, Lu:2017meb, Luchinsky:2020fdf, Olamaei:2020bvw, Rahmani:2020pol, Shi:2017dto, Shi:2019fph, Shi:2020qde, Wang:2017azm, Wang:2017mqp, Xiao:2017dly, Xiao:2017udy, Xing:2018lre, Yao:2018ifh, Zhang:2018llc, Zhao:2018mrg}. (See also the review paper \cite{aliyev2022}).
The electromagnetic moments and form factors represent promising tools for understanding the inner structure of baryons. The multipole moments are connected with the spatial charge and current distributions of baryons. For this reason, the determination of the multipole moments of baryons represents an important issue in obtaining useful insight on the internal structure of baryons.
In the present work, we study the multipole moments of the $J^P=\frac{3}{2}^+$ doubly heavy baryons within the light-cone QCD sum rules (LCSR) framework. It should be noted that the magnetic dipole moments of the doubly heavy decuplet baryons and multipole moments of the decuplet baryons within the same framework are studied in \cite{ozdem2020} and \cite{Aliev:2009pd}, respectively.
The structure of this work is as follows. In section II we derive the light-cone sum rules for the relevant electromagnetic form factors. Section III is devoted to the analysis of the form factors obtained in section II.

\section{LCSR for Multipole Moments of $J^P=\frac{3}{2}^+$ Doubly Heavy Baryons}
    For obtaining the LCSR for multipole moments of doubly heavy baryons with $J^P=\frac{3}{2}^+$, we consider the following correlation function.
\begin{equation}
    \Pi_{\mu\alpha\nu}(p,q) = i^2\int d^4x\int d^4ye^{ipx+iqy}\mel{0}{\mathcal{T}\left\{\eta_\mu(x)j^{em}_\alpha(y)\bar{\eta}_\nu(0)\right\}}{0}
    \label{sec2:correl0}
\end{equation}
where $\eta_\mu$ is the interpolating current of the doubly heavy baryon (here and in all next discussions, we call the $J^P=\frac{3}{2}^+$ doubly heavy baryon as doubly heavy baryon for shortness.), $j^{em}_\alpha$ is the electromagnetic current, $p$ and $q$ are four momenta of the final baryon and the electromagnetic current respectively.
The interpolating current of the doubly heavy spin-$3/2$ baryon is
\begin{equation}
    \eta_\mu = \dfrac{N}{\sqrt{3}}\varepsilon^{abc}\Big\{\left(q^{aT}C\gamma_\mu Q^b\right)Q^{\prime c}+\left(q^{aT}C\gamma_\mu Q^{\prime b}\right)Q^{c}+\left(Q^{aT}C\gamma_\mu Q^{\prime b}\right)q^{c}\Big\}
    \label{sec2:interpolating}
\end{equation}
where $q$ means the light quark($u,d$ or $s$), $a,b$ and $c$ are the color indices, $C$ is the charge conjugation operator. $N$ is the normalization factor which is equal to $\sqrt{2}$ when $Q\neq Q^{\prime}$ and $1$ when $Q=Q^{\prime}$. Introducing the electromagnetic background field of a plane wave
\begin{equation}
    F_{\mu\nu} = i(\varepsilon_\mu q_\nu - \varepsilon_\nu q_\mu)e^{iqx}
    \label{sec2:emwave}
\end{equation}
the correlation function given in Eq. (\ref{sec2:correl0}) can be written as
\begin{equation}
    \varepsilon^\alpha\Pi_{\mu\alpha\nu}(p,q)=i\int d^4xe^{ipx}\mel{0}{\mathcal{T}\Big\{\eta_\mu(x)\bar{\eta}_\nu(0)\Big\}}{0}_F
    \label{sec2:correl1}
\end{equation}
In Eq. (\ref{sec2:correl1}), the subscript $F$ means that all condensates are calculated in the presence of the background field. Expanding Eq. (\ref{sec2:correl1}) in the power of background field and retaining only the linear terms in $F_{\mu\nu}$ which corresponds to radiation of the photon, we can get the correlation function given by Eq. (\ref{sec2:correl0}). (Technical details of the background field method can be found in \cite{Ball:2002ps} and \cite{Novikov:1984ecy}.)
According to the standard strategy of QCD sum rules, the correlation function should be calculated in two different kinematical domains. If $p^2,p^{\prime2}>0$ then the correlation function given in Eq. (\ref{sec2:correl1}) can be written in terms of hadrons. In this kinematical region the hadronic representation of the correlation function is obtained as
\begin{equation}
    \varepsilon^\alpha\Pi_{\mu\alpha\nu}(p,q) = \dfrac{\varepsilon^\alpha \mel{0}{\eta_\mu(0)}{B^\ast(p)}\mel{B^\ast(p)}{j^{em}_\alpha}{B^\ast(p+q)}\mel{B^\ast(p+q)}{\bar{\eta}_\nu(0)}{0}}{\left[(p+q)^2-m_1^2\right]\left[p^2-m_2^2\right]}+\cdots
    \label{sec2:correl2}
\end{equation}
where $\cdots$ denotes higher state and continuum contributions. The matrix element $\mel{0}{\eta_\mu}{B^\ast(p)}$ describes the coupling of $\eta_\mu$ current to the $B^\ast$ and defined as 
\begin{equation}
    \mel{0}{\eta_\mu}{B^\ast(p)} = \lambda u_\mu(p)
    \label{sec2:mel0}
\end{equation}
where $\lambda$ is the residue and $u_\mu(p)$ is the Rarita-Schwinger spinor for the $J=\frac{3}{2}$ particle. 
The matrix element $\mel{B^\ast(p^\prime)}{j^{em}_\alpha}{B^\ast(p)}$ is parametrized in terms of four form factors as follows
\begin{equation}
    \begin{aligned}
        \mel{B^\ast(p^\prime)}{j^{em}_\alpha}{B^\ast(p)} &= -\bar{u}_{\beta^\prime}(p^\prime)\Big\{g^{\beta^\prime\beta}\left[\gamma_\alpha F_1\left(Q^2\right)+i\dfrac{\sigma_{\alpha\rho}q^\rho}{m_1+m_2}F_2\left(Q^2\right)\right]\\
        &-\dfrac{2q^{\beta^\prime}q^\beta}{\left(m_1+m_2\right)^2}\left[\gamma_\alpha F_3\left(Q^2\right)+i\dfrac{\sigma_{\alpha\rho}q^\rho}{m_1+m_2}F_4\left(Q^2\right)\right]\Big\}u_\beta(p)
    \end{aligned}
    \label{sec2:mel1}
\end{equation}After employing the Wick theorem, both massive and
massless quark propagators appear in the pre
Putting Eqs.(\ref{sec2:mel0}) and (\ref{sec2:mel1}) into Eq. (\ref{sec2:correl2}) and performing summation over spins of $J^P = \frac{3}{2}^+$ baryon with the help of the formula
\begin{equation}
    \sum u^{(s)}_\alpha(p)\bar{u}^{(s)}_\beta(p) = (\slashed{p}+m)\left\{-g_{\alpha\beta}+\dfrac{1}{3}\gamma_\alpha\gamma_\beta-\dfrac{2}{3}\dfrac{p_\alpha p_\beta}{m^2}-\dfrac{1}{3}\left(p_\alpha\gamma_\beta-p_\beta\gamma_\alpha\right)\right\}
    \label{sec2:rarita-schwinger}
\end{equation}
in principle, we can obtain the hadronic part of the correlation function. At this point, we face with a problem: Interpolating current $\eta_\mu$ interacts not only with the $J=\frac{3}{2}$ states but also $J=\frac{1}{2}$ states. The matrix element of $\eta_\mu$ between the vacuum and $J=\frac{1}{2}$ one particle state is defined as 
\begin{equation}
    \mel{0}{\eta_\mu}{B(p,s=1/2)} = \left(Ap_\mu+B\gamma_\mu\right)u(p)
    \label{sec2:spin1-2_mel}
\end{equation}
 From this expression, it follows that the contributions of $J=\frac{1}{2}$ baryons are either proportional to $p_\mu$,$p^\prime_\nu$, or $\gamma_\mu$ at left or $\gamma_\nu$ at the right. Therefore, in order to discard the contributions of $J=\frac{1}{2}$ states, we will neglect the structures proportional to $p_\mu$, $p^\prime_\nu$ and structures with $\gamma_\mu$ at the left and $\gamma_\nu$ at the right.
Another problem we face is that not all Lorentz structures are independent. To overcome this problem we order the Dirac matrices in a specific order. In the present work, we choose Dirac matrices in the following order $\gamma_\mu\slashed{p}\slashed{\varepsilon}\slashed{q}\gamma_\nu$. Taking into account these observations, the correlation function can be represented as 
\begin{equation}
    \begin{aligned}
        \varepsilon^\alpha\Pi_{\mu\alpha\nu}(p,q) &= \Big\{g_{\mu\nu}(p.\varepsilon)\slashed{p}\Pi_1 + g_{\mu\nu}(p.\varepsilon)\slashed{q}\Pi_2\\
        &+(\varepsilon.p) q^{\mu}q^{\nu}\Pi_3+\slashed{p}\slashed{q}(\varepsilon.p) q^{\mu}q^{\nu}\Pi_4 +\dots\Big\}
    \end{aligned}
    \label{sec2:correl_hadronic}
\end{equation}

where 
\begin{equation}
    \begin{aligned}
        \Pi_1 &= -\dfrac{2\lambda_1\lambda_2}{\left((p+q)^2-m_1^2\right)\left(p^2-m_2^2\right)}F_1\\
        \Pi_2 &= -\dfrac{\lambda_1\lambda_2}{\left((p+q)^2-m_1^2\right)\left(p^2-m_2^2\right)}F_2\\
        \Pi_3 &= -\dfrac{\lambda_1\lambda_2}{\left((p+q)^2-m_1^2\right)\left(p^2-m_2^2\right)}\dfrac{m_2}{(m_1+m_2)^2}F_3\\
        \Pi_4 &= -\dfrac{\lambda_1\lambda_2}{\left((p+q)^2-m_1^2\right)\left(p^2-m_2^2\right)} \dfrac{2}{(m_1+m_2)^3}F_4
    \end{aligned}
    \label{sec2:ffcorrels}
\end{equation}
The QCD side of the correlation function can be calculated in the deep Euclidean region, where $p^2\ll0$ and $p^{\prime2}\ll0$. The correlation function in this domain can be expressed in terms of photon distribution amplitudes (DAs). To obtain it, the expression of the interpolating current is inserted into the correlation function in Eq. (\ref{sec2:correl1}). After employing the Wick theorem, the correlation function can be expressed in terms of both massive and massless quark propagators as:
\begin{equation}
    \begin{aligned}
        \varepsilon^\alpha\Pi_{\mu\alpha\nu}(p,q) &= i\dfrac{N^2}{3}\epsilon^{abc}\epsilon^{a^\prime b^\prime c^\prime}\int d^4xe^{ipx} \\
        &\langle0\vert\Big\{ -S_Q^{cb^\prime}\gamma_\nu\widetilde{S}_{Q^\prime}^{aa^\prime}\gamma_\mu S_q^{bc^\prime} - S_Q^{ca^\prime}\gamma_\nu\widetilde{S}_{Q^\prime}^{bb^\prime}\gamma_\mu S_q^{ac^\prime} - S_{Q^\prime}^{ca^\prime}\gamma_\nu\widetilde{S}_{Q}^{bb^\prime}\gamma_\mu S_q^{ac^\prime}\\
        &- S_{Q^\prime}^{cb^\prime}\gamma_\nu\widetilde{S}_{q}^{aa^\prime}\gamma_\mu S_Q^{bc^\prime} - S_{q}^{ca^\prime}\gamma_\nu\widetilde{S}_{Q^\prime}^{bb^\prime}\gamma_\mu S_Q^{ac^\prime} - S_{q}^{cb^\prime}\gamma_\nu\widetilde{S}_{Q}^{aa^\prime}\gamma_\mu S_{Q^\prime}^{bc^\prime} \\
        &-S_{Q^\prime}^{cc^\prime}\Tr\left[S_{Q}^{ba^\prime}\gamma_\nu\widetilde{S}_{q}^{ab^\prime}\gamma_\mu\right] - S_{q}^{cc^\prime}\Tr\left[S_{Q^\prime}^{ba^\prime}\gamma_\nu\widetilde{S}_{q}^{ab^\prime}\gamma_\mu\right]- S_{Q}^{cc^\prime}\Tr\left[S_{q}^{ba^\prime}\gamma_\nu\widetilde{S}_{Q^\prime}^{ab^\prime}\gamma_\mu\right]\Big\}\vert0\rangle_F
    \end{aligned}
\end{equation}
where 
\begin{equation}
    \begin{aligned}
        S_q(x) &= S^{free}_q(x)-ig_s\dfrac{1}{16\pi^2x^2}\int_0^1du\big\{\bar{u}\slashed{x}\sigma_{\alpha\beta}+u\sigma_{\alpha\beta}\slashed{x}\big\}G^{\alpha\beta}(ux)\\
        S_Q(x) &= S^{free}_Q(x)-ig_s\dfrac{m_Q}{16\pi^2}\int_0^1duG^{\alpha\beta}(ux)\\
        &\left\{\sigma_{\alpha\beta}K_0\left(m_Q\sqrt{-x^2}\right)+\left((\bar{u}\slashed{x}\sigma_{\alpha\beta}+u\sigma_{\alpha\beta}\slashed{x})\dfrac{K_1\left(m_Q\sqrt{-x^2}\right)}{\sqrt{-x^2}}\right)\right\}
    \end{aligned}
    \label{sec2:quarkprops}
\end{equation}
where the free propagators of the heavy and the light quarks are 
\begin{equation}
    \begin{aligned}
        S^{free}_q(x) &= \dfrac{i\slashed{x}}{2\pi^2x^4}\\
        S^{free}_Q(x) &= \dfrac{m_Q^2}{4\pi^2}\left[\dfrac{K_1\left(m_Q\sqrt{-x^2}\right)}{\sqrt{-x^2}}+\dfrac{i\slashed{x}}{\left(\sqrt{-x^2}\right)^2}K_2\left(m_Q\sqrt{-x^2}\right)\right]\\
    \end{aligned}
\end{equation}
where $K_i\left(m_Q\sqrt{-x^2}\right)$ are the modified Bessel function of the second kind, $G_{\mu\nu}$ is the background gluonic field strength tensor.

The correlation function contains two kinds of contributions: Perturbative and non-perturbative. Perturbative contributions correspond to the case when the photon interacts with quarks perturbatively, and the second one corresponds to the case where the photon interacts with quarks at large distances.
The perturbative contributions are obtained by replacing the propagator of the quark that interacts with the photon perturbatively with
\begin{equation}
    S_{\alpha\beta}^{ab} \rightarrow \left\{\int d^4y S^{free}(x-y)\slashed{A}S^{free}(y)\right\}_{\alpha\beta}^{ab}
\end{equation}
For the calculation of the non-perturbative part, the light quark propagator is replaced with
\begin{equation}
    S_q(x) \rightarrow -\dfrac{1}{4}\bar{q}(x)\Gamma_i q(0) \Gamma_i
    \label{sec2:pertreplacement}
\end{equation}
where $\Gamma_i$ is the full set of Dirac matrices. In this case matrix elements $\mel{0}{\bar{q}(x)\Gamma_iq(0)}{0}_F$ and $\mel{0}{\bar{q}(x)\Gamma_iG_{\mu\nu}q(0)}{0}_F$ appear, which are expressed in terms of photon DAs\cite{Ball:2002ps}. Photon DAs are the main non-perturbative input parameters.\\
The form factors $F_i$ can be obtained by matching the coefficients of the relevant Lorentz structures in hadronic parts onto their corresponding operator product expansion (OPE) part of the correlation functions. In addition, we have used the quark-hadron duality ansatz to eliminate the contributions of the continuum and excited states in the hadronic dispersion relation. After applying this ansatz and performing Borel transformations on variables $-p^2$ and $-(p+q)^2$, we get the desired sum rules for the form factors $F_i(q^2)$ at $q^2=0$ point.
\begin{equation}
    \lambda_1\lambda_2F_i(0) e^{-m_1^2/M_1^2}e^{-m_2^2/M_2^2}=\int ds_1\int ds_2 e^{-s_1/M_1^2}e^{-s_2/M_2^2}\rho_i(s_1,s_2)
    \label{sec2:ff_spectral0}
\end{equation}
where $\lambda_1$ and $\lambda_2$, $m_1$ and $m_2$ are the residues and the masses of the initial and final baryons respectively.
Introducing new variables $s_1 = \dfrac{su}{u_0}$ and $s_2 = \dfrac{s\bar{u}}{\bar{u}_0}$, where $u_0 = \dfrac{M_2^2}{M_1^2+M_2^2}$, Eq. (\ref{sec2:ff_spectral0}) can be written as
\begin{equation}
    \lambda_1\lambda_2F_i(0) e^{-m_1^2/M_1^2}e^{-m_2^2/M_2^2}=\int_{(m_Q+m_{Q^\prime}+m_q)^2}^{s_0} dse^{-s/M^2}\rho_i\left(s\right)
    \label{sec2:ff_spectral1}
\end{equation}
where $M^2 = \dfrac{M_1^2M_2^2}{M_1^2+M_2^2}$ and $\displaystyle\rho_i(s) = \dfrac{s}{u_0\bar{u}_0}\int_0^1du\rho_i\left(\dfrac{su}{u_0},\dfrac{s\bar{u}}{\bar{u}_0}\right)$. This amounts to subtracting the contributions of higher states and continuum by keeping only the triangular region $u_0s_1+\bar{u}_0s_2\le s_0$ in the $(s_1, s_2)$ plane. Since the initial and the final baryons are same in our case, $\lambda_1 = \lambda_2=\lambda$, $m_1=m_2=m$, and we can also set $M_1^2 = M_2^2$ leading to $u_0=\dfrac{1}{2}$, and $M_1^2=M_2^2=2M^2$. From Eq. \ref{sec2:ff_spectral1} it follows that for the determination of $F_i(0)$, the residues of $J=\frac{3}{2}$ baryons are needed. These residues are calculated in \cite{Aliev:2012iv} within the QCD sum rules method, which we will use in the numerical calculations.
As we already noted, from the experimental point of view the multipole form factors(moments) are more suitable. Therefore, relations between $F_i(0)$ and multipole form factors are needed. These relations for an arbitrary spin baryon form factors are obtained in \cite{Lorce:2009br}. For a real photon, these relations are
\begin{equation}
    \begin{aligned}
        F_{2k+1}(0) &= \sum_{l=0}^kC_{n-l}^{n-k}(-1)^{k-l}G_{E_{2l}}(0)\\
        F_{2k+2}(0) &= \sum_{l=0}^kC_{n-l}^{n-k}(-1)^{k-l}\left[G_{M_{2l+1}}(0)-G_{E_{2l}}(0)\right]\\
    \end{aligned}
    \label{sec2:ff_relations_general}
\end{equation}
where $C_n^k = \begin{cases}
    \dfrac{n!}{k!(n-k)!}&\text{for}\;\;n\geq k\geq0\\
    0&\text{otherwise}
\end{cases}$ and $j=n+\frac{1}{2}$ where $j$ is the spin of the particle. Taking into account these definitions, one can easily find the following relations between two sets of form factors at the $q^2=0$ point.
\begin{equation}
    \begin{aligned}
        F_1(0)&=G_{E_0}(0)\\
        F_2(0)&=G_{M_1}(0)-G_{E_0}(0)\\
        F_3(0)&=-G_{E_0}(0)+G_{E_2}(0)\\
        F_4(0)&=-G_{M_1}(0)+G_{E_0}(0)+G_{M_3}(0)-G_{E_2}(0)
    \end{aligned}
    \label{sec2:ff_relations}
\end{equation}
\section{Numerical Calculations}
    This section is devoted to the analysis of the sum rules for the form factors. For the values, the input parameters appearing in the sum rules are \cite{braun2016,Ball:2002ps,belyaev1983,navas2024,shifman1979}
    \begin{equation}
        \begin{aligned}
            m_c &= (1.4\pm0.01)\;\text{GeV},\\
            m_b &= (4.8\pm0.015)\;\text{GeV},\\
            f_{3\gamma} &= -0.0039\;\text{GeV}^2,\\
            \chi &= (3.15\pm0.10)\;\text{GeV}^2,\\
            \ev{\bar{q}q} &= (-0.24\pm0.001)^3\;\text{GeV}^3,\\
            \omega^V_\gamma &= 3.8 \pm 1.8,\\
            \omega^A_\gamma &= -2.1 \pm 1.0
        \end{aligned}
        \label{parameters}
    \end{equation}
    The masses of the spin-$3/2$ baryons are calculated in \cite{Aliev:2012iv,brown2014}. We present their masses in Table \ref{tab:baryonmasses}.

\begin{table}[!ht]
    \centering
    \begin{tabular}{|l|c|c|}
        \hline
         Baryon & Lattice\cite{brown2014} & QCDSR\cite{Aliev:2012iv} \\
         \hline
         $\Xi^\ast_{cc}$ & $3.692$ GeV & $3.69$ GeV \\
         \hline
         $\Xi^\ast_{bc}$ & $6.985$ GeV & $7.25$ GeV \\
         \hline
         $\Xi^\ast_{bb}$ & $10.178$ GeV & $10.4$ GeV \\
         \hline
    \end{tabular}
    \caption{Baryon Masses}
    \label{tab:baryonmasses}
\end{table}
The sum rules for the form factors contain also two auxiliary parameters: the continuum threshold $s_0$, and the Borel mass parameter $M^2$. The working region of the $s_0$ is determined from two-point sum rules \cite{Aliev:2012iv} and given in Table \ref{tab:workingregions}.
The working region of $M^2$ depends on two requirements: $M^2$ should be large enough to guarantee the dominance of the leading twist and small enough to suppress the higher state and continuum contributions. The working regions of the Borel mass parameters satisfying both of these conditions are also given in Table \ref{tab:workingregions}.
\begin{table}[!ht]
    \centering
    \begin{tabular}{|l|c|c|}
        \hline
         Baryon & $M^2\; (\text{GeV}^2)$ & $s_0\; (\text{GeV}^2)$  \\
         \hline
         $\Xi_{cc}$ & $3-6$ & $19\pm 1$\\
         \hline
         $\Xi_{bc}$ & $6-9$ & $59\pm 1$\\
         \hline
         $\Xi_{bb}$ & $9-12$ & $121\pm 2$\\
         \hline
    \end{tabular}
    \caption{Working regions of $M^2$ and $s_0$}
    \label{tab:workingregions}
\end{table}

\begin{figure}[!ht]
    \centering
    \subfloat[]{%
        \includegraphics{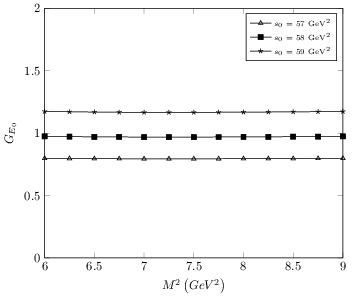}%
    }
    \subfloat[]{%
        \includegraphics{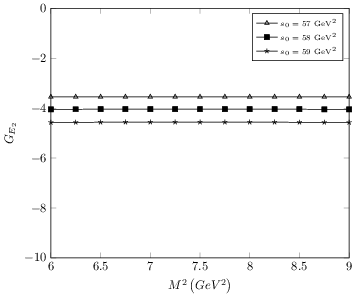}%
    }\\
    \subfloat[]{%
        \includegraphics{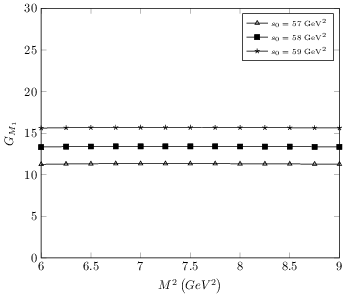}%
    }
    \subfloat[]{%
        \includegraphics{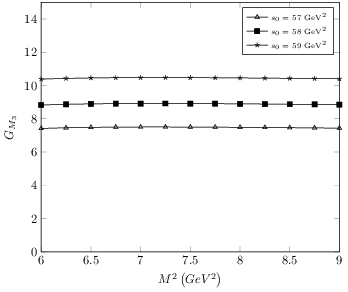}%
    }
    \caption[$M^2$ dependence for $\Xi^{\ast+}_{bc}$]{\textbf{a)} $M^2$ dependence of the $G_{E_0}$ for $\Xi^{\ast+}_{bc}$ baryon with different $s_0$ values, \textbf{b)} same as in \textbf{a)} but for the $G_{E_2}$ , \textbf{c)} same as in \textbf{a)} but for the $G_{M_1}$, \textbf{d)} same as in \textbf{a)} but for the $G_{M_3}$. All values are presented in figures are in natural units}
    \label{xisbcpgs}
\end{figure}

\begin{figure}[!ht]
    \centering
    \subfloat[]{%
        \includegraphics{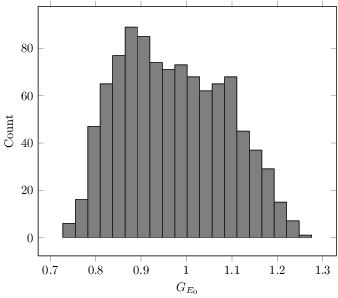}%
    }
    \subfloat[]{%
        \includegraphics{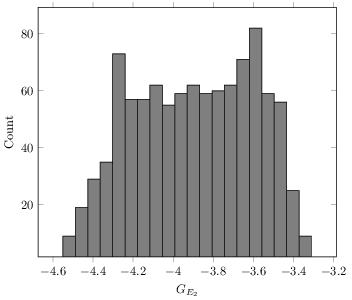}%
    }\\
    \subfloat[]{%
        \includegraphics{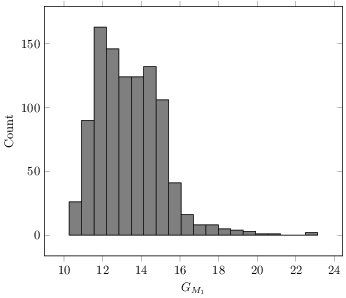}%
    }
    \subfloat[]{%
        \includegraphics{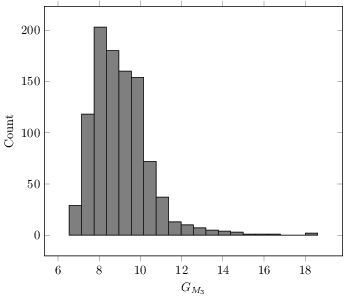}%
    }
    \caption[Histograms]{Histograms for multipole moments of the $\Xi^{\ast+}_{bc}$ baryon}
    \label{hists}
\end{figure}

In Fig. \ref{xisbcpgs}, we present the dependencies of the multipole moments of $\Xi_{bc}^{\ast+}$ on $M^2$ at different fixed values of $s_0$ using the central values of the input parameters. From these figures, we observe that the multipole moments of $J^P=\frac{3}{2}^+$ doubly heavy baryons exhibit good stability with respect to the variation of $M^2$ in its working region. To obtain all uncertainties, including the uncertainty arising from the input parameters as well as the Borel mass and the continuum threshold, we follow the procedure proposed in \cite{Leinweber:1995fn}: 1000 sets of random values chosen in the parameter space given in Eq. (\ref{parameters}) and Table \ref{tab:workingregions}. The histograms for the $\Xi_{bc}^+$ baryon are given in Fig \ref{hists}. Our results are given in Table \ref{tab:results} (in the natural magneton for multipole moments).

\begin{table}[!ht]
    \centering
    \begin{tabular}{|c|c|c|c|}
        \hline
        Baryon & $G_{E_2}$ & $G_{M_1}$ & $G_{M_3}$\\
        \hline
        $\Xi_{cc}^{\ast++}$&$0.15\pm0.13$&$7.93\pm1.16$&$6.23\pm0.96$\\
        \hline
        $\Xi_{cc}^{\ast+}$&$2.46\pm0.38$&$-0.10\pm0.13$&$1.22\pm0.29$\\
        \hline
        $\Xi_{bb}^{\ast0}$&$-9.70\pm1.06$&$20.32\pm2.77$&$11.42\pm1.90$\\
        \hline
        $\Xi_{bb}^{\ast-}$&$3.10\pm0.25$&$-13.67\pm2.22$&$-9.58\pm1.91$\\
        \hline
        $\Xi_{bc}^{\ast+}$&$-4.04\pm0.23$&$14.17\pm3.14$&$9.64\pm3.06$\\
        \hline
        $\Xi_{bc}^{\ast0}$&$1.83\pm0.13$&$-2.92\pm0.30$&$-1.34\pm0.23$\\
        \hline
    \end{tabular}
    \caption{Electromagnetic multipole results in natural magneton}
    \label{tab:results}
\end{table}
The magnetic dipole moments $G_{M_1}$ of doubly heavy baryons are studied in various approaches. In Table \ref{tab:spin32results}, we present our results on magnetic dipole moment (in the nuclear magneton units) and the results of other works. 

\begin{table}[!ht]
    \centering
    \resizebox{\textwidth}{!}{%
    \begin{tabular}{|c|c|c|c|c|c|c|}
        \hline
         Works & $\Xi^{\ast ++}_{cc}$ & $\Xi^{\ast +}_{cc}$ & $\Xi^{\ast 0}_{bb}$ & $\Xi^{\ast -}_{bb}$ & $\Xi^{\ast +}_{bc}$ & $\Xi^{\ast 0}_{bc}$ \\
         \hline
          This Work& $2.08\pm0.34$ & $-0.03\pm0.03$ & $1.84\pm0.25$ & $-1.24\pm0.20$ & $1.80\pm0.23$ & $-0.39\pm0.04$\\
          \hline
          Bag \cite{bagmodel-32}& 2.001 & 0.163 & 0.916 & -0.652 & 1.414 & -0.257 \\
         \hline
         exBag \cite{simonis2018}& 2.35 & -0.178 & 1.40 & -0.880 & 1.88 & -0.534 \\
         \hline
         NRQM \cite{NRQM-32} & $2.67^{+0.27}_{-0.15}$ & $-0.311^{+0.052}_{-0.130}$ & $1.87^{+0.27}_{-0.13}$ & $-1.11^{+0.06}_{-0.14}$ & $2.27^{+0.27}_{-0.14}$ & $-0.712^{+0.059}_{-0.133}$ \\
         \hline
         EMS \cite{effectiveqmscheme-32} & $2.4344 \pm 0.0033$ & $-0.0846 \pm 0.0025$ & $1.5897 \pm 0.0016$ & $-0.9809 \pm  0.0008$ & $2.0131 \pm 0.0020$ & $-0.5315 \pm 0.0012$ \\
         \hline
         NRQM \cite{NRQM2-32} & 2.676 & -0.165 & 1.767 & -1.074 & 2.222 & -0.620 \\
         \hline 
         QM \cite{QM-32,chiralpt-32} & 2.61 & -0.18 & 1.73 & -1.06 & 2.17 & -0.62 \\
             \hline
         $\chi_{PT}$\cite{chiralpt-32} & 1.72 & -0.09 & 0.63 & -0.79 & 1.12 & -0.40 \\
         \hline
         LCSR\cite{ozdem2020} & 2.94 & -0.67 & 2.30 & -1.39 & 2.63 & -0.96\\
         \hline
    \end{tabular}}
    \caption{Magnetic dipole moments in nuclear magneton units}
    \label{tab:spin32results}
\end{table}

From Table\ref{tab:spin32results}, we observed that
\begin{itemize}
    \item Central value for the magnetic dipole moment of $\Xi_{cc}^{\ast++}$ is in good agreement with the result of the Bag model\cite{bagmodel-32}, and exBag model\cite{simonis2018} especially and approximately 50\% less than the result of the \cite{ozdem2020}.
    \item Our result for $\Xi_{bb}^{\ast0}$ (central value) is very close to NRQM\cite{NRQM-32,NRQM2-32}, QM\cite{QM-32,chiralpt-32} and considerably differ from ones obtained in Bag model\cite{bagmodel-32} and $\chi_{PT}$\cite{chiralpt-32}
    \item For the $\Xi_{bb}^{\ast-}$ case, our result is close to the results of NRQM\cite{NRQM-32}, LCSR\cite{ozdem2020} and differ from the results of other approaches.
    \item For $\Xi_{bc}^{\ast+}$, our findings are very close to exBag\cite{simonis2018},EMS\cite{effectiveqmscheme-32} and notable different than others.
    \item Finally, for $\Xi_{bc}^{\ast0}$, our results are close to the results of exBag\cite{simonis2018}, $\chi_{PT}$\cite{chiralpt-32} and substantially different than other approaches.
\end{itemize}

As we already noted that the magnetic dipole moments of doubly heavy baryons with $J^P=\frac{3}{2}^+$ within the same framework are calculated in \cite{ozdem2020}. From the comparison, we see that our results differ mainly from the ones obtained in \cite{ozdem2020}. The difference can be explained by two reasons: 
\begin{itemize}
    \item Difference in the values of the input parameters
    \item Our results on the spectral density do not coincide with \cite{ozdem2020}
\end{itemize}

\section{Conclusion}
In the present work, we calculated the multipole moments of doubly heavy baryons in the framework of light cone QCD sum rules. Measurement of multipole moments can play an essential role in understanding the inner structure of the doubly heavy baryons.
We also perform a comparison of our findings of magnetic dipole moments with existing literature results. We obtained that there are considerable differences among predictions of different approaches.
Further improvements in the various approaches will make predictions more accurate. More theoretical studies are needed in this direction.

\bibliographystyle{apsrev4-1}
\bibliography{ref}

\newpage

\section{Figures}
\pgfplotsset{width=10cm,compat=1.18}


\newpage
\appendices
\section{Correlation Functions}
In this appendix, we present the expressions of the correlation functions corresponding to the Lorentz structures $(\varepsilon p)\slashed{p}g^{\mu\nu}$, $(\varepsilon p)\slashed{q}g^{\mu\nu}$, $(\varepsilon p)q^\mu q^\nu$ and $(\varepsilon p)\slashed{p}\slashed{q}q^\mu q^\nu$ correspondingly.
\begin{align}
        \Pi_1 &= \dfrac{1}{8\pi^4}\Big\{ 2e_Qm_Q^2\left(I^3_{-1,0,1}-I^3_{-1,1,1}-I^3_{0,0,1}\right) + 2e_{Q^\prime}m_{Q^\prime}^2\left(I^3_{0,-1,1}-I^3_{1,-1,1}-I^3_{0,0,1}\right)\\
		\nonumber&+e_Q\left(I^3_{0,1,2}-I^3_{1,1,2}-I^3_{0,2,2}\right)+e_{Q^\prime}\left(I^3_{1,0,2}-I^3_{1,1,2}-I^3_{2,0,2}\right)\\
        \nonumber&+4(e_q+e_Q+e_{Q^\prime})m_Qm_{Q^\prime}\left(I^3_{0,0,1}-I^3_{1,0,1}-I^3_{0,1,1}\right)\\
		\nonumber&+(3e_q+2e_Q+2e_{Q^\prime})I^3_{0,0,2}+3(e_q+e_Q+e_{Q^\prime})\left(I^3_{0,0,2}-I^3_{0,1,2}-I^3_{1,0,2}\right)\Big\}\\
        \Pi_2 &= \dfrac{1}{144\pi^4}\Bigg\{
        3 \pi^2 \mathbb{A}\left(u_0\right) e_q \ev{\overline{q}q}\left((m_Q^3+m_{Q^\prime}^3)I^1_{0,0} + m_Qm_{Q^\prime}\left(m_QI^1_{-1,1}+m_{Q^\prime}I^1_{1,-1}\right)\right)\\
		\nonumber&+2 \pi^2 e_q (m_Q+m_{Q^\prime}) \ev{\overline{q}q}I^2_{0,0,0}\\
		\nonumber&\mathcal{F}_1\left((4-6\alpha_g-8\alpha_{\bar{q}})\mathcal{S}-\alpha_g \mathcal{T}_1 - (1-\alpha_g+2\alpha_{\bar{q}}) \mathcal{T}_3 + (-1 + 2\alpha_{\bar{q}})\mathcal{T}_4\right)\\
		\nonumber&+2\pi^2 \ev{\overline{q}q} (e_Qm_{Q^\prime} + e_{Q^\prime}m_Q) \mathcal{F}_1\left((4-6\alpha_g-8\alpha_{\bar{q}})\mathcal{S}_{\gamma}-(1-\alpha_g+2\alpha_{\bar{q}}) \mathcal{T}_{4\gamma}\right)I^2_{0,0,0}\\
		\nonumber&+2\pi^2e_qf_{3\gamma}I^2_{0,0,1}\\
		\nonumber&\left[-\mathcal{F}_3\left(\dfrac{\alpha_g+\alpha_{\bar{q}}-u_0}{\alpha_g}(3\mathcal{A}+2\mathcal{V})\right)+\mathcal{F}_4\left(\dfrac{\alpha_g+\alpha_{\bar{q}}-u_0}{\alpha_g}(3\mathcal{A}+2\mathcal{V})\right)-\mathcal{F}_1\left((3\mathcal{A}+2\mathcal{V})\right)\right]\\
		\nonumber&+2\pi^2e_q\ev{\overline{q}q}\mathcal{F}_1\left(\mathcal{S}-\widetilde{\mathcal{S}}-\mathcal{T}_1-\mathcal{T}_2-\mathcal{T}_3-\mathcal{T}_4\right)\left(m_{Q^\prime}I^2_{-1,1,0}+m_QI^2_{1,-1,0}\right)\\
		\nonumber&+2\pi^2\ev{\overline{q}q}\mathcal{F}_1\left(\mathcal{T}_{4\gamma}-\alpha_g\mathcal{S}_{\gamma}\right)\left(e_{Q^\prime}m_{Q^\prime}I^2_{-1,1,0}+e_Qm_QI^2_{1,-1,0}\right)\\
		\nonumber&+6e_q\pi^2\ev{\overline{q}q}\left(m_{Q^\prime}I^2_{0,1,0}+m_QI^2_{1,0,0}\right)\mathbb{A}\left(u_0\right)\\
		\nonumber&+9e_qI^2_{1,1,2}-12\pi^2e_q\chi\ev{\overline{q}q}\varphi_\gamma\left(u_0\right)\left(m_{Q^\prime}I^2_{0,1,1}+m_QI^2_{1,0,1}\right)+2\pi^2e_qf_{3\gamma}\psi^a\left(u_0\right)I^2_{1,1,1}\\
		\nonumber&+8\pi^2e_qf_{3\gamma}\mathcal{U}\left[\psi^v(u)\right]\left(2m_Qm_{Q^\prime}I^2_{0,0,0}+2I^2_{1,1,1}+m_Q^2I^2_{1,0,0}+m_{Q^\prime}^2I^2_{0,1,0}\right)\\
		\nonumber&+9m_Qm_{Q^\prime}\Big[-\left(e_QI^3_{-1,0,1}+e_{Q^\prime}I^3_{0,-1,1}\right)+\left(e_QI^3_{-1,1,1}+e_{Q^\prime}I^3_{1,-1,1}\right)\\
		\nonumber&+\left(e_Q+e_{Q^\prime}\right)\left(5I^3_{0,0,1}-4I^3_{1,0,1}-4I^3_{0,1,1}\right)\Big]\\
		\nonumber&+9\Big[-\left(e_QI^3_{0,1,2}+e_{Q^\prime}I^3_{1,0,2}\right)\\
		\nonumber&+2\left(e_QI^3_{0,2,2}+e_{Q^\prime}I^3_{2,0,2}\right)\\
		\nonumber&+3\left(e_Q+e_{Q^\prime}\right)\left(I^3_{1,1,2}-I^3_{2,1,2}-I^3_{1,2,2}\right)\Big]\\
		\nonumber&+18\Big[e_Qm_Q^2I^3_{-1,1,1}+e_{Q^\prime}m_{Q^\prime}^2I^3_{1,-1,1}\\
		\nonumber&- e_Qm_Q^2I^3_{-1,2,1} - e_{Q^\prime}m_{Q^\prime}^2I^3_{2,-1,1}\\
		\nonumber&- e_Qm_Q^2I^3_{0,1,1} - e_{Q^\prime}m_{Q^\prime}^2I^3_{1,0,1}\Big]\\
		\nonumber&+9e_q\Big[2m_Qm_{Q^\prime}\left(I^3_{0,0,1}-2I^3_{0,1,1}-2I^3_{1,0,1}\right) - \left(I^3_{1,1,2}+3I^3_{1,2,2}+3I^3_{2,1,2}\right)\Big]\Bigg\}\\
        \Pi_3 &= \dfrac{1}{144\pi^4}\Bigg\{-48 \pi^2 e_q\ev{\overline{q}q}m_Qm_{Q^\prime} \mathcal{U}\left[h_{\gamma}(u)(2u-1)\right]I^2_{0,0,0}\\
		\nonumber&-12 \pi^2 e_q\ev{\overline{q}q} \mathcal{U}\left[h_{\gamma}(u)(8u-1)\right] I^2_{1,1,0}\\
		\nonumber&+6e_q\pi^2\ev{\overline{q}q}\mathbb{A}\left(u_0\right)\left(m_{Q^\prime}^2I^1_{0,1}+m_Q^2I^1_{1,0}\right)\\
		\nonumber&+4e_q\pi^2\ev{\overline{q}q} I^2_{0,0,0}\\
		\nonumber&\mathcal{F}_1\Bigg(\left(\alpha_g^2+u_0(1-2\alpha_{\bar{q}})^2+3\alpha_g(\alpha_{\bar{q}}-1)\right)\mathcal{S}\\
		\nonumber&-\left(\alpha_g^2+2(\alpha_{\bar{q}}-4)\alpha_{\bar{q}}+\alpha_g(3\alpha_{\bar{q}}-5)+\dfrac{7}{2}\right)\widetilde{\mathcal{S}}\\
		\nonumber&-\left(\alpha_g^2+\alpha_g(\alpha_{\bar{q}}-5)-3\alpha_{\bar{q}}+\dfrac{3}{2}\right)\mathcal{T}_2 + 2\left(1 + \alpha_g^2+\alpha_{\bar{q}}(2\alpha_{\bar{q}}-3)+\alpha_g(3 \alpha_{\bar{q}}-2)\right)\mathcal{T}_3\\
		\nonumber&+\left(\alpha_{\bar{q}}+\alpha_g+(\alpha_g+\alpha_{\bar{q}}-1)-u_0\right)\mathcal{T}_4\Bigg)\\
		\nonumber&-4\pi^2e_qf_{3\gamma}\left(m_Q+m_{Q^\prime}\right)I^2_{0,0,0}\mathcal{F}_1\left((1+\alpha_g-2\alpha_{\bar{q}})\mathcal{A}-(8\alpha_g+8\alpha_{\bar{q}}-4)\mathcal{V}\right)\\
		\nonumber&+18\pi^2e_q\ev{\overline{q}q}\mathbb{A}\left(u_0\right)I^2_{1,1,0}\\
		\nonumber&-4\pi^2\ev{\overline{q}q}\left(e_{Q^\prime}I^2_{0,1,0}+e_QI^2_{1,0,0}\right)\\
		\nonumber&\mathcal{F}_1\left(\left(\alpha_g^2+3\alpha_g(\alpha_{\bar{q}}-1)+u_0(2\alpha_{\bar{q}}-1)^2\right)\mathcal{S}_{\gamma} -2\left((\alpha_g-1)^2 + \alpha_{\bar{q}}(2\alpha_{\bar{q}}+3\alpha_g-3)\right)\mathcal{T}_{4\gamma}\right)\\
		\nonumber&+4\pi^2e_qf_{3\gamma}\left(m_QI^2_{0,-1,0}+m_{Q^\prime}I^2_{-1,0,0}-4\left(m_Q+m_{Q^\prime}\right)I^2_{0,0,0}\right)\mathcal{F}_1\left(\alpha_g\mathcal{V}\right)\\
		\nonumber&-8\pi^2e_q\ev{\overline{q}q}\left(I^2_{0,1,0}+I^2_{1,0,0}\right)\mathcal{F}_2\left((1+23v)\mathcal{T}_1\right)\\
		\nonumber&+32\pi^2e_q\ev{\overline{q}q}\left(10I^2_{1,1,0}-\left(I^2_{0,1,0}+I^2_{1,0,0}\right)+5\left(I^2_{0,2,0}+I^2_{2,0,0}\right)\right)\mathcal{F}_2\left(\mathcal{T}_1\right)\\
		\nonumber&-136\pi^2e_q\ev{\overline{q}q}\mathcal{F}_2\left((1+v)\mathcal{T}_2\right)\left(I^2_{0,1,0}+I^2_{1,0,0}\right)\\
		\nonumber&+40\pi^2e_q\ev{\overline{q}q}\mathcal{F}_2\left((3+2v)\mathcal{T}_2\right)\left(I^2_{2,0,0}+I^2_{1,1,0}+I^2_{0,2,0}\right)\\
		\nonumber&+8e_q\pi^2\ev{\overline{q}q}\mathcal{F}_2\left((2v-1)\mathcal{T}_3\right)\left(5I^2_{2,0,0}-7I^2_{1,0,0}+10I^2_{1,1,0}-7I^2_{0,1,0}+5I^2_{0,2,0}\right)\\
		\nonumber&+8\pi^2e_q\ev{\overline{q}q}\mathcal{F}_2\left((2v-3)\mathcal{T}_4\right)\left(I^2_{0,1,0}+I^2_{1,0,0}\right)\\
		\nonumber&-8\pi^2\ev{\overline{q}q}\mathcal{F}_2\left((2v-1)\mathcal{T}_{4\gamma}\right)\\
		\nonumber&\left(7(e_{Q^\prime}I^2_{0,1,0}+e_QI^2_{1,0,0}) - 5(e_{Q^\prime}I^2_{0,2,0}+e_QI^2_{2,0,0}) - 5(e_{Q^\prime}+e_Q)I^2_{1,1,0}\right)\\
		\nonumber&-30\pi^2e_qf_{3\gamma}\psi^a\left(u_0\right)\left(m_QI^2_{1,0,0}+m_{Q^\prime}I^2_{0,1,0}\right)\\
		\nonumber&-24\pi^2e_q\ev{\overline{q}q}\chi\varphi_{\gamma}\left(u_0\right)I^2_{1,1,1}\\
		\nonumber&-18\left(e_{Q^\prime}m_{Q^\prime} \left(I^3_{1,-1,1}-I^3_{2,-1,1}\right)+e_Qm_Q\left(I^3_{-1,1,1}-I^3_{-1,2,1}\right)\right)\\
		\nonumber&-36(e_q+e_Q+e_{Q^\prime})\left(m_QI^3_{0,2,1}-\left(m_Q+m_{Q^\prime}\right)I^3_{1,1,1}+m_{Q^\prime}I^3_{2,0,1}\right)\\
		\nonumber&-27\left(e_Qm_{Q^\prime}+e_{Q^\prime}m_Q\right)I^3_{0,0,1}\\
		\nonumber&+9\left(\left(7e_{Q^\prime}+e_q\right)m_Q + 3e_Q\left(2 m_Q + m_{Q^\prime}\right)\right)I^3_{0,1,1}\\
		\nonumber&+9\left(\left(7e_Q+e_q\right)m_{Q^\prime} + 3e_{Q^\prime}\left(2 m_{Q^\prime} + m_Q\right)\right)I^3_{1,0,1}\\
		\nonumber&+8\pi^2e_q\ev{\overline{q}q}\mathcal{F}_2\left(4\mathcal{T}_1+(3+2v)\mathcal{T}_2+(2v-1)\mathcal{T}_3\right)\left(m_Q^2+m_{Q^\prime}^2\right)I^1_{2,2}\\
		\nonumber&+8\pi^2\ev{\overline{q}q}\mathcal{F}_2\left((2v-1)\mathcal{T}_{4\gamma}\right)\left(m_Q^2e_Q+m_{Q^\prime}^2e_{Q^\prime}\right)I^1_{2,2}\\
		\nonumber&-8\pi^2e_q\ev{\overline{q}q}\mathcal{F}_2\left(4\mathcal{T}_1 + (3+2v)\mathcal{T}_2 + (2v-1)\mathcal{T}_3\right)\left(m_Q^2I^1_{1,2}+m_{Q^\prime}^2I^1_{2,1}\right)\\
		\nonumber&-8\pi^2\ev{\overline{q}q}\mathcal{F}_2\left((2v-1)\mathcal{T}_{4\gamma}\right)\left(m_Q^2e_QI^1_{1,2}+m_{Q^\prime}^2e_{Q^\prime}I^1_{2,1}\right)\\
		\nonumber&+8\pi^2e_q\ev{\overline{q}q}\mathcal{F}_2\left(4\mathcal{T}_1 + (3+2v)\mathcal{T}_2 + (2v-1)\mathcal{T}_3\right)\left(m_Q^2I^1_{1,-1}+m_{Q^\prime}^2I^1_{-1,1}\right)\\
		\nonumber&-8\pi^2e_q\ev{\overline{q}q}\mathcal{F}_2\left(4\mathcal{T}_1 + (3+2v)\mathcal{T}_2 + (2v-1)\mathcal{T}_3\right)\left(m_{Q^\prime}^2I^1_{0,-1}+m_Q^2I^1_{-1,0}\right)\\
		\nonumber&+16\pi^2e_q\ev{\overline{q}q}\mathcal{F}_2\left(3\mathcal{T}_1+3\mathcal{T}_2-\mathcal{T}_3-\mathcal{T}_4\right)m_Qm_{Q^\prime}I^1_{-1,-1}\\
		\nonumber&-16\pi^2\ev{\overline{q}q}\mathcal{F}_2\left(\mathcal{T}_{4\gamma}\right)m_Qm_{Q^\prime}\left(e_QI^1_{0,-1}+e_{Q^\prime}I^1_{-1,0}\right)\\
		\nonumber&-8\pi^2\ev{\overline{q}q}\mathcal{F}_2\left((2v-1)\mathcal{T}_{4\gamma}\right)\left(e_Qm_{Q^\prime}2I^1_{0,-1}+e_{Q^\prime}m_Q^2I^1_{-1,0}\right)\Bigg\}\\
        \Pi_4 &= \dfrac{e_q\ev{\overline{q}q}}{18\pi^2}\Bigg\{2\mathcal{F}_2\left(\mathcal{T}_4(\alpha_i)-\mathcal{T}_2(\alpha_i)\right)I^1_{0,0}+3\mathcal{U}\left[\left(2u-1\right)h_\gamma(u)\right]I^1_{1,1}\Bigg\}
\end{align}

Where 
\begin{equation}
    \begin{aligned}
        I^1_{m,n} & = \int_0^{s_0}dse^{-s/M^2}\int_0^1dx x^m (1-x)^n \delta\left(s-\dfrac{m_Q^2}{1-x}-\dfrac{m_{Q^\prime}^2}{x}\right)\\
        I^2_{m,n,t} & = \int_0^{s_0}dse^{-s/M^2}\int_0^sd\alpha\int_0^1dx x^m (1-x)^n (s-\alpha)^t \delta\left(\alpha-\dfrac{m_Q^2}{1-x}-\dfrac{m_{Q^\prime}^2}{x}\right)\\
        I^3_{m,n,t} & = \int_0^{s_0}dse^{-s/M^2}\int_0^sd\alpha\int_0^1dx\int_0^{1-x}dy x^m y^n (s-\alpha)^t \delta\left(\alpha-\dfrac{m_Q^2}{x}-\dfrac{m_{Q^\prime}^2}{y}\right)\\
    \end{aligned}
\end{equation}

and

\begin{equation}
    \begin{aligned}
        \mathcal{F}_1(DA) &=\int \mathcal{D}\alpha_i \dfrac{DA(\alpha_i)}{\alpha_g^2} \theta\left(\alpha_g + \alpha_{\bar{q}} - u_0\right)\theta\left(-\alpha_{\bar{q}} + u_0\right)\\
		\mathcal{F}_2(DA) &= \int \mathcal{D}\alpha_i\int_0^1dv DA(\alpha_i) \theta\left(\alpha_g + \alpha_{\bar{q}} - \alpha_g v - \frac{1}{2})\right)\\
		\mathcal{F}_3(DA)&= \int \mathcal{D}\alpha_i\dfrac{DA(\alpha_i)}{\alpha_g^2}\delta\left(\dfrac{\alpha_g+\alpha_{\bar{q}}-u_0}{\alpha_g}\right)\theta\left(u_0-\alpha_{\bar{q}}\right)\\
		\mathcal{F}_4(DA)&= \int \mathcal{D}\alpha_i\dfrac{DA(\alpha_i)}{\alpha_g^2}\delta\left(\dfrac{\alpha_{\bar{q}}-u_0}{\alpha_g}\right)\theta\left(\alpha_{\bar{q}}+\alpha_g-u_0\right)\\
        \mathcal{U}\left[f(u)\right] &= \int_0^1duf(u)\theta\left(u-u_0\right)
    \end{aligned}
\end{equation}

where 

\begin{equation}
    \mathcal{D}\alpha_i = d\alpha_gd\alpha_qd\alpha_{\bar{q}}\delta\left(1-\alpha_g-\alpha_q-\alpha_{\bar{q}}\right)
\end{equation}

\end{document}